# Meta-analysis of scRNA-seq data for choroidal endothelial cells in dry Age-related Macular Degeneration

**Kyle M. Veksler[1,2,*], Levi Dong[1,2,*], Timothy A. Blenkinsop[1,2,@], Aurelian Radu[1,@]**

[*] These authors had equal contribution

[1] Dept. Stem Cell Biology and Regenerative Medicine, Icahn School of Medicine at Mount Sinai, One Gustave L Levy Place, New York, NY 10029, USA;

[2] Dept. of Ophthalmology, Icahn School of Medicine at Mount Sinai, One Gustave L Levy Place, New York, NY 10029, USA

[@] Corresponding authors: aurelian.radu@mssm.edu, timothy.blenkinsop@mssm.edu

**ABSTRACT.** The mechanisms that lead to dry Age-related Macular Degeneration are largely unelucidated, which prevents the introduction of effective therapies. Experimental support exists in the literature for the hypothesis that choroidal endothelial cell (ChEC) dysfunction precedes the loss of macular retinal pigmented epithelial (RPE), which may be only a secondary consequence of inadequate blood supply. If so, interventions at the level of ChEC could constitute an under investigated therapeutic strategy. Datasets regarding the transcriptional changes in early or intermediate dry AMD are publicly available, but for some some of them the information about ChECs have not been analyzed, or not analyzed using the most powerful and recent software tools. We present here new data generated by our bioinformatics analysis of these datasets. The main new finding is that angiogenesis is initiated in dry AMD, as it is in wet AMD. However, contrary to wet AMD, in dry AMD angiogenesis fails to execute, and therefore the blood supply that supports the RPE becomes gradually insufficient, leading to their dysfunctionality and death. The data support a unitary hypothesis of the origin / initiation / etiology of both dry and wet AMD, namely that both are initiated by ChEC dysfunction - either insufficient / abortive angiogenesis in dry AMD, or excessive angiogenesis in wet AMD. Pathway analysis also reveals as perturbed Notch and TNF signaling, endothelial to mesenchymal transition (EndoMT), mitochondria, "fluid shear stress", "osteoclast differentiation" and "calcification/osteoporosis". Overall, the new data provide a rationale for experimental studies, to validate and further characterize these perturbations, and investigate strategies to correct them.

**Abbreviations:**

AMD - age-related macular degeneration
eAMD - early AMD
iAMD - intermediate AND
GA - geographic atrophy
nGA - nascent GS
ChECs - choroidal endothelial cells

CC - choriocapillaris
RPEs - retinal pigment epithelial cells
DEGs - differentially expressed genes

## INTRODUCTION

Dry (non-proliferative or non-exudative) age-related macular degeneration (AMD) is the leading cause of vision loss in older adults. It is characterized by progressive dysfunction of the macula and eventual loss of central vision (Fleckenstein et al., 2021). The number of individuals affected with some form of AMD globally in 2020 was estimated to be 196 million and is projected to be 288 million in 2040 (Fleckenstein et al., 2024).

The disease has multiple variations and stages, beginning with early AMD (eAMD) followed by intermediate AMD (iAMD), differentiated by drusen deposition and subtle retinal changes. The clinical hallmarks of eAMD are early basal laminar deposits of basement-membrane-like material, including collagen, laminin, fibronectin, and other extracellular matrix components with no pigmentary abnormalities. These deposits exist as medium-sized drusen spots (64-124 µm in diameter) or multiple small-sized drusen (<64 µm in diameter). The hallmarks of iAMD are large drusen deposits (>125 µm in diameter), RPE migration from their original attachment at the Bruch's membrane into more inner layers of the retina, and/or pigmentary changes (Fleckenstein et al., 2021; Fleckenstein et al., 2024). Late stages of the disease manifest as either dry AMD or wet AMD. Late dry AMD, also known as geographic atrophy (GA), is characterized by loss of retinal pigment epithelial (RPE) and photoreceptors within a radius of 500 µm from the foveal center. Wet AMD, referred to as proliferative or exudative AMD, is characterized by excessive proliferation of blood vessels (choroidal neovascularization) (Fleckenstein et al., 2021).

**Available therapeutic interventions.** Therapies for wet AMD are effective and are widely used. They are based on anti-VEGF antibodies that reduce endothelial cell growth (Yan et al., 2025). In contrast, therapies for dry AMD have very limited efficacy. The therapies are based on the discovery that complement components have a major genetic contribution to dry AMD. Two complement inhibitors have been recently approved by the FDA for clinical use in GA patients. The first, approved in August 2023, Pegcetacoplan, is a peptide that acts as a complement C3 inhibitor (Nadeem et al., 2023). Pegcetacoplan slows down AMD progression by 14% per year. The second, Izervay, is an RNA aptamer that binds complement protein C5 and inhibits its cleavage, blocking downstream MAC formation. Izervay decreases disease progression by 18 - 22% per year (Yan et al., 2025). JNJ-1887 (in clinical trials) is an adeno-associated virus (AAV) gene therapy agent developed for geographic atrophy that increases the expression of CD59/protectin, a protein in the final steps of the MAC cascade, that stops pores from forming (Yan et al., 2025). The Membrane Attack Complex (MAC) is the final step of the complement cascade, which leads to the formation of pores in pathogens' cell membranes, causing them to lyse, but inadvertently can cause lysis of endogenous cells. Studies support the view that MAC depositions form in eAMD, potentially before further disease progression (Arya et al., 2018). Another agent in clinical trials is Luminate (risuteganib), a synthetic RGD peptide for iAMD, which competes with integrins (Shaw et al., 2020). Alternatively, cell therapies for replacing the lost RPE cells are undergoing clinical trials. RG6501 and EyeCyte-RPE consist of allogenic RPEs, RPESC-RPE-4W is derived from cadaver donor eyes (Rao et al., 2025), while ASP7317 uses RPE cells derived from human embryonic stem cells (Yan et al., 2025).

**RPE versus the choroidal ChECs as the optimal therapeutic target.** Consensus in the AMD research community on the cellular chain of events that leads to dry AMD is not resolved . Even the type of cells where the disease is initiated is under debate. The most studied candidate consists of the RPE, because they are responsible for the clear pathological markers of the disease: the drusen deposits and their loss precedes vision loss. Drusen is made up of lipids and proteins generated by imperfect processing by the RPE of remnants of the photoreceptors due to the RPE role in photoreceptor survival, outer segment phagocytosis, and subretinal environment maintenance.

Moreover, the loss of macular RPE is the direct, easily visible, and consequential endpoint of the pathology. This is the reason why the vast majority of the research effort is concentrated on the RPE. However, this may not be the optimal therapeutic strategy. RPE malfunction may not be cell autonomous but instead is initiated by a defective blood supply provided by the choroidal blood vessels. The choriocapillaris (CC) is the capillary bed that supplies the outer retina. A key observation is that the retina has extremely high oxygen consumption and oxidative stress burden, with the RPE acting as a key protective/maintenance tissue in this high-demand environment (Strauss 2005; Simo et al., 2010; Joyal et al., 2017). Based on this possibility, even a mild decline in oxygen and nutrient delivery, caused by a decline in the retinal blood vessels, could produce significant and irreversible damage to the RPEs. ChECs may represent a therapeutic target, with potential for efficacy in the early disease, before irreversible RPE loss occurs, and even disease prevention. ChEC dysfunction may precede RPE changes and can be the driver of AMD pathogenesis.

We review below the evidence supporting that (a) changes in the choroidal vasculature occur in dry AMD, (b) they are correlated with AMD markers of progression, and (c) the choroidal changes precede changes in RPEs.

**Choroidal vasculature occur degenerates in dry AMD.** Spraul et al. (1996) examined 19 deceased eyes of non-neovascular AMD patients and 40 age-matched controls using microscopy, finding that the number of choroidal veins decreased, with a difference of 3.5 choroidal veins/mm compared to 5.7 choroidal veins/mm in controls ($p < 0.001$).

Mullins et al. (2014) demonstrated increased deposition of membrane attack complex (MAC) markers in the choroid of aged eyes ($p < 0.001$) and even greater MAC deposition in AMD donor eyes ($p < 0.05$) using enzyme-linked immunosorbent assay analysis (MAC is on the cells); individuals homozygous with the high AMD risk allele had reduced choroidal thickness by 23.6% ($p < 0.05$) over individuals homozygous for the low risk allele (Arya et al., 2018). Lutty et al. (2020) quantified a reduction in submacular vascular area in 7 early AMD eyes compared to 7 age-matched controls (60.1 ± 10.4% vs 78.1 ± 3.25%, $p < 0.0001$).

Edwards et al. (2023) examined the eyes of three siblings with AMD and controls, finding that the submacular vascular area is significantly reduced with geographic atrophy, declining from 78.1% ± 3.2% in controls to 20.6% ± 5% in GA ($p < 0.0005$), alongside significant reductions in CC lumen diameter (16.5 ± 2.25 μm vs. 9.6 ± 2.9 μm, $p < 0.001$).

AMD eyes show choriocapillaris loss occurring in early/intermediate AMD and increasing toward advanced stages of the disease (Seddon et al., 2016).

CC degenerates markedly in dry AMD, and there is a linear relationship between the loss of RPE and CC in GA (McLeod et al., 2009).

Cell-free choriocapillaris domains (ghost vessels) exist in atrophic age-related macular degeneration (Mullins et al., 2024).

**Changes in the choroidal vasculature are correlated with markers of AMD**. Mullins et al. (2011) looked at the eyes of 21 early AMD and 24 age-matched controls using fluorescence microscopy, observing that choroid density decreased to some extent with greater drusen (correl. coeff. = $r^2$ = 0.22, $p < 0.01$) and that ghost vessels, a lower or non-functioning blood vessel remnants, were also strongly associated with drusen presence ($r^2$ = 0.57, $p < 0.001$) in eAMD.

Borrelli et al. (2020) showed that macular drusen-free regions of intermediate AMD patients exhibit reduced photoreceptor density and abnormal choroidal flow deficits, and a significant inverse correlation of photoreceptor density exists with choroidal flow deficit %. Flow deficits are defined as a lack of detectable blood movement.

Similarly, in the “drusen-free” regions, the Ellipsoid Zone normalized reflectivity, a measurement used to quantify retinal health, was inversely correlated with the choroidal flow deficit % ($\rho = -0.340$, $p = 0.020$). In control eyes, the Ellipsoid Zone normalized reflectivity was not associated with the choroidal variables ($\rho = -0.075$ and $p = 0.668$ for flow deficit % and $\rho = 0.007$ and $p = 0.969$ for flow deficit average size).

The important implication of these data is that reduction of choroidal flow and reduction in photoreceptor density or in the EZ normalized reflectivity are not caused by drusen as a common causal factor. This observation is compatible with the hypothesis that choroidal changes precede drusen formation.

Importantly, no correlation was observed in control eyes, which means that the correlation appears as a consequence of pathological processes in AMD.

Nassisi et al. (2019) studied 33 eyes of 23 patients using spectral domain OCT and spectral domain OCTA, once at baseline and a follow-up at least 12 months later. The authors examined two regions: the para-atrophic region, defined as the innermost 500 μm-wide ring immediately surrounding the edge of the atrophic lesion, and the peri-atrophic region, a concentric 500 μm-wide ring outside the para-atrophic region. They demonstrated that in para-atrophic regions, CC flow voids (FV) at baseline correlated with yearly GA growth rate in ($R = 0.579$, $p < 0.001$). The difference between flow voids in the para- and peri-atrophy choroidal regions correlated with yearly GA growth rate ($R = 0.681$, $p < 0.001$). Therefore, choroidal degeneration at earlier stages is correlated with later GA enlargement.

Structural OCT and OCTA were used to examine patients with intermediate AMD at baseline and again after at least 12 months. In the eyes that progressed to complete retinal pigment epithelial and outer retinal atrophy, choroidal flow density was significantly greater than in those that did not ($p < 0.0001$) (Corvi et al., 2021).

A greater inner choroid flow deficit and a lower scotopic microperimetric retinal sensitivity at baseline were predictors of progression to iRORA (incomplete retinal pigment epithelial and outer retinal atrophy) in eyes with intermediate AMD (Corradetti et al., 2021).

Vidal-Oliver et al. (2024) longitudinally looked at 46 non-advanced AMD patients (presence of drusen; subretinal drusenoid deposits; hyperreflective foci; and no evidence of active or previous exudation, central atrophy, or scar tissue) and 26 controls (patients over the age of 50 without symptoms). They found no relationship in non-advanced AMD between choroid thickness and the development of advanced AMD. Choroid thickness is correlated with capillary density (Fujiwara et al., 2016). The inverse correlation is compatible with the hypothesis that either abnormal blood flow causes reduced photoreceptor density, possibly due to RPE malfunction, or the opposite, reduced photoreceptor density reduces the demand for blood supply and leads by a feedback mechanism to a reduction in blood flow.

In animal models, acute RPE ablation by sodium iodate or surgical/mechanical debridement leads to secondary choriocapillaris atrophy, supporting a causal sequence opposite to that favored in this study. However, these findings should be interpreted cautiously, because such models involve acute, artificial RPE injury and may not faithfully reflect the chronic, multifactorial progression of human AMD (Korte et al., 1984; Upadhyay & Bonilha, 2025)

These data do not establish a cause-and-effect relation – they are compatible with either of the two possibilities (1) EC dysfunction leads to less oxygen and nutrients and loss of RPE, or (2) loss or damage of RPE leads to less demand for oxygen and nutrients, leading to atrophy of choroidal.

**Studies supporting the view that changes in the choriocapillaris are a causal factor for dry AMD.** The above studies do not establish a cause-and-effect relation – they are compatible with either of the two possibilities (1) EC dysfunction leads to less oxygen and nutrients and loss of RPE, or (2) loss or damage of RPE leads to less demand for oxygen and nutrients, leading to atrophy of choroidal. The following studies support more explicitly the view that dysfunction of the choroidal is a cause of RPE loss.

Biesemeier et al. (2014) graded postmortem eyes in five stages of progressing disease and examined retina sections by electron microscopy. In the AMD retinas, the areas covered by choroidal lumina were reduced by half compared with controls, and the number of fenestrations in the remaining endothelia was greatly decreased. Choroid was always more damaged in stage II compared with the RPE, suggesting that loss of choroid defects precedes RPE defects.

Nattagh et al. (2020) observed 12 eyes of dry AMD subjects using OCTA (Optical Coherence Tomography Angiography) at baseline and at follow-ups of 7-16 months. The choroidal flow deficit at baseline had an odds ratio (OR) of 1.33 as a predictor of developing GA ($p = 0.017$). They conclude that choroidal Flow Void % likely reflects hypoperfusion and is pathogenically important in GA progression.

Romano et al. (2024) used swept source OCTA to examine 13 eyes with early AMD and 51 eyes with intermediate AMD, with at least 2 imaging points over at least 18 months. They reported that progression from intermediate AMD to GA significantly correlated with baseline intercapillary-flow deficit % and with baseline subfoveal choroid thickness, supporting the view that FD precedes GA development

Tiosano et al. (2021) examined 34 patients with intermediate AMD longitudinally at one point and after 12 months, using Swept-Source OCT and OCTA. In 25 of these patients, flow deficit % increased significantly, while intermediate AMD, evaluated structurally, did not progress. Their interpretation is that CC changes may appear before structural changes that reflect AMD progression. However, they did not prove that iAMD progression in the 25 patients occurred after 12 months.

Two years of treatment with monthly injections of anti-VEGF medication was associated with a 59% increase in risk of GA development, suggesting that reduced EC proliferation could be a cause of GA (Grunwald et al., 2014).

Greig et al. (2024) examined the relationship between choroidal loss and the development of nascent GA (nGA) evaluated as drusen volume, in 105 intermediate AMD eyes using swept source OCTA, in patients who did not have nGA or late AMD at the initial visit. Imaging was repeated every 6-months, with each patient making at least 2 visits. Choroidal flow deficit % and drusen volume measurements were determined for the visit prior to nGA development or the second-to-last visit if nGA did not develop. Regions identified to have high choroidal flow deficit % in an earlier measurement subsequently developed nGA in a future imaging session, evaluated by drusen volume. Notably, no association between choroidal flow deficit % and the subsequent development of nGA was found at the global level (across the macula). A strong association was found only at the local, superpixel level, indicating precise, early focal choroidal dysfunction in future susceptible areas. This study strongly indicates that choroidal flow deficits precede and could cause of nGA. Area under the curve (AUC) graphs were used to assess the ability of flow deficit or drusen to predict the development of nGA. Choroidal flow deficits were slightly, but statistically significantly superior to drusen volume (0.84 vs 0.80 AU). The addition of drusen to choroidal flow deficit did not improve the AUC compared with flow deficit only, which suggests that flow deficits are the cause, and drusen is a closely associated effect of choroidal flow deficits. The AUC value is, however, too low to be clinically useful for prediction.

Li et al. (2022) showed that leukemia inhibitory factor, a mitogen for choroidal endothelial cells, reduces choroidal flow deficits and protects against retinal atrophy in a GA mouse model.

Choriocapillaris loss was observed in early AMD, even in areas of intact retinal pigment epithelium (Sohn et al., 2019).

Together the summarized evidence suggests a causal role of choroid flow deficits in dry AMD. However this evidence is circumstantial and more focused experiments are needed to confirm a causal relationship.

In principle, longitudinal clinical imaging studies can be used to investigate a causal relation between choroidal flow and AMD progression. Future studies are needed to examine the temporal changes of localized choroidal flow deficits prior to nGA development, to understand the timeframe of choroidal loss.

Multiple longitudinal data points can be used as input for Granger causality software packages to determine if the choroidal changes precede in time the pathological markers of AMD progression. To our knowledge, no study has applied this strategy to dry AMD. It remains possible that both processes are caused by another unknown factor, with choroidal changes  are a faster consequence than RPE loss.

A definitive answer may be obtained only by treating AMD patients with agents that reduce the defect in the choroid and determine if the disease is alleviated. This is not possible so far, because such agents are not known. Discovery of such agents is difficult because the nature and the relative importance of the cellular and molecular processes that occur in the choroid in AMD are so far less known. Systematic progress requires studies that will clarify these processes, which can then be followed by a rational search for therapeutic agents.

Future studies are needed to examine the temporal changes in such localized choroid flow deficits prior to nGA development to examine this notion further and understand the timeframe of choroid loss, which could be achieved through evaluating longitudinal cohorts seen over multiple follow-up visits (Grieg et al., 2024).

**scRNA-seq studies of chECs in dry AMD.** The data generated by a study on 3 early AMD cases, 1 atrophic, and 5 controls have been deposited in the public repository GEO (Gene Expression Omnibus - https://www.ncbi.nlm.nih.gov/geo/) under the reference no. GSE203499. The data are public since Jun 23, 2022. The field "Citation" does not list any publication associated with this dataset, and we did not find any publications that analyzed the ECs included. Therefore we conducted a bioinformatic analysis of the EC data, referred to below as Dataset 1 in the Results section.

Another study reported scRNA-seq data from RPE/choroid cells from one intermediate AMD and one unaffected subject (Collin et al., 2023). Assessment of the statistical significance of the differences cannot be done, due to the low number of samples.

A study by Voigt et al. (2022), analyzed samples from 10 control and 9 early atrophic (non-neovascular or 'dry') AMD donors. CD31 enrichment of ChECs from the dissociated cell samples was used prior to scRNA-seq. The conclusion of the study was that in the choriocapillaris, no genes yielded a false discovery rate (FDR) of less than 0.05 after correction for multiple hypotheses.

Another study (Zauhar et al., 2022), from 2 early AMD donors and 1 control, was deposited in GEO under the reference no. GSE188280. Statistical significance cannot be assessed due to the low number of cases.

A higher number of samples, 6 AMD and 7 control donors, was analyzed by (Orozco et al., 2023). However, the 6 AMD cases contain both geographic atrophy and neovascular cases, in a non-specified proportion. Given the very different characteristics of chECs in the two forms of the disease, it would be difficult to assess the significance of these data for dry AMD. Moreover, the expression analysis lists DEGs for multiple cell types, but not ChECs.

Another dataset, GSE230348, analyzed data from 7 normal controls, 1 early AMD, 7 intermediate, and 3 atrophic AMD cases.  The data are public since Jan 18, 2024. The field "Citation" does not list any publication associated with this dataset, and we did not find any publications that analyzed the ECs included. Our analysis, referred to below as Dataset 2, is presented in the Results section.

The most recent study (Wang et al., 2026, GSE262151), referred to below as Dataset 3, analyzed 14 postmortem eyes of nine individuals, of which six were normal controls, 2 early/intermediate, and one advanced AMD. The samples consisted of retina and choroid single nuclei. The authors noticed that nuclei capture from RPE and choroid was initially hampered by excess pigment interfering with single-cell microfluidics. They modified the procedure, enabling much higher nuclei yields. The comparison of control and AMD eyes showed 2-56 differentially expressed genes per cell type. The reported statistical significance of DEGs for ChECs is extremely high, e.g., p-adj = 1.72E-32. It is extremely improbable that such values can be obtained with the low number of cases available and after adjusting for multiple testing. A plausible explanation is that the significance was computed using as input all single nuclei, which vastly overestimates the statistical significance. If large groups of nuclei in the comparison belong to a single donor, they do not represent independent measurements, which is the basic assumption of the applied statistical test. The test must be applied to pseudobulk values, which combine all the nuclei derived from the same donor into a single number. Our analysis of the data, referred to below as Dataset 3, is presented in the Results section. Therefore in total we analyzed three separate unpublished datasets.

## METHODS

**Public datasets and sample preparation.** Publicly available single-cell RNA-sequencing (scRNA-seq) datasets profiling human retina and/or RPE/choroid were obtained from the NCBI Gene Expression Omnibus (GEO, https://www.ncbi.nlm.nih.gov/geo/). The datasets used were: GSE203499 (Dataset 1), GSE230348 (Dataset 2) and GSE262151 (Dataset 3).

For the datasets that contained both retina and RPE/choroid compartments, only the RPE/choroid-relevant libraries were retained for downstream ChEC-focused analyses.

All selected samples were standardized into a common sparse matrix input format comprising a Matrix Market count matrix (matrix.mtx) and associated feature and barcode files (features.tsv, barcodes.tsv). A registry file (dataset_registry.csv) was constructed to define the processing inputs for each library and to provide harmonized metadata fields (dataset identifier, sample identifier, patient/donor label, and condition label). When datasets contained distinct biological replicates represented as separate libraries (e.g., separate eyes or independently processed regions), these were retained as independent samples in the registry rather than collapsed at the donor level, preserving the correct replicate structure for downstream aggregation and statistical testing.

**Single-cell processing and endothelial enrichment (Seurat).** Each library was imported into Seurat (Hao et al., 2023) as an independent object. Cell barcodes were renamed using a sample-specific prefix to prevent barcode collisions during cross-study merging. Per-cell QC metrics were computed, including mitochondrial transcript fraction (percent.mt, MT- genes) and ribosomal fraction (percent.ribo, RPL/RPS genes). Cells were retained if they met all of the following thresholds: nFeature_RNA ≥ 200, nCount_RNA ≥ 3, nCount_RNA ≤ 5×10^5, percent.mt < 10, and percent.ribo < 20. Doublets were detected per sample using scDblFinder (expected doublet rate 0.05) and removed prior to downstream analysis. Samples then underwent iterative clustering-based cleanup. Briefly, data were normalized, highly variable genes were selected (2,000 features), and PCA was performed, followed by UMAP visualization (dims=30) and clustering (resolution=0.3). Cluster markers were identified using positive markers only (min.pct=0.6, logFC threshold=1), and clusters with mixed or ambiguous identity relative to curated marker panels were removed prior to reclustering (up to four iterations).

Endothelial cells were enriched using a two-stage strategy. EC-enriched clusters were preferentially retained based on marker enrichment. When cluster-level assignment was insufficient, a strict cell-level gate was applied, requiring an EC module score ≥80th percentile and expression of ≥3 EC markers, along with a veto of ≤1 non-EC marker hit.

**Integration, pseudobulk aggregation, and differential expression.** Cleaned objects were merged for global analysis and integrated using Harmony to reduce sample-associated effects.

For the identification of the Differentially Expressed Genes (DEGs), the pseudobulk data were obtained by merging the data from all the cells from each donor. The data were analyzed using the R package Deseq2 (Love et al., 2014). The Benjamini–Hochberg correction for multiple testing was applied. Key intermediate plots and tables were exported throughout, and session information was recorded for reproducibility.

Heat maps were generated using the R package pheatmap (Kolde et al., 2025). Identification of the pathways enriched in the DEG sets was done using the QIAGEN IngenuityPathway Analysis (IPA) (accessed January 19, 2026) (Krämer et al., 2014). For the Fast Gene Set Enrichment Analysis, we used the R package fgsea (Korotkevich et al., 2021), which is based on the original GSEA package (Subramanian et al., 2005). The Bioconductor ROntoTools was used as described in (Ansari et al., 2017; Voichita et al., 2024). The R package PathfindR for pathway analysis was used as described in (Ulgen et al., 2019).

## RESULTS AND DISCUSSION

### DIFFERENTIALLY EXPRESSED GENES (DEGs)

We focused on early and intermediate dry AMD, rather than the advanced GA stage. The reason is that, from a therapeutic point of view, the most effective interventions would be applied in the early/intermediate stage, rather than in the advanced, presumably irreversible stage. Additionally, for the purpose of understanding the mechanisms responsible for the disease, the early stages are most informative and easier to interpret, compared with the late stages, in which multiple secondary processes can obscure the initiating causes.

Datasets 1 and Dataset 2. The two datasets subjected to DESeq2 analysis yielded sufficiently large numbers of DEGs to be analyzed by pathway analysis packages, as described in the next section.

For Dataset 3**,** we obtained EC clusters for 6 controls and 2 early/intermediate AMD donors. The analysis yielded 2 statistically significant DEGs, MYO1E (myosin 1E) and ARHGEF28, both upregulated in the AMD samples. MYO1E does not have a known direct connection with ECs or with AMD. MYOE1 plays a role in TGFB receptors' internalization and in their transport to the endosomal compartment (Chung et al., 2019). ARHGEF28 is involved in cyclic-stretch-induced perpendicular reorientation of endothelial cells (Abiko et al., 2015). This process is relevant for arterial and possibly arteriolar ECs.

### PATHWAY ANALYSIS

We used four analysis packages, in order to increase the confidence that the commonly identified pathways play roles in AMD: Ingenuity Pathway Analysis, PathfindR**,** ROntoTools, and fGSEA.

**Ingenuity Pathway Analysis (IPA)** is a widely used analysis and visualization commercial platform built on an expert-curated knowledge base (QIAGEN Inc., Redwood City, CA, (Krämer et al., 2015)).

Dataset 1. EC clusters were obtained from 5 early AMD and 4 control cases. The pseudobulk counts were analyzed by DESeq2 (log2Fc cutoff +/- 0.38 (equivalent to +/-130% linear), pAdj cutoff = 0.2); 40 genes up, 5 down. (Due to the relatively low number of DEGs and taking into account the exploratory nature of the inquiry, a less stringent p-adj cutoff was used, 0.2. This offers, for the worst-case scenario — the genes that are close to the threshold — a false positives ratio of 1 in 5, which is still acceptable to guide discovery and further inquiries).

The DEGs were used as input for pathway analysis.

The graphical summary of the analysis is presented in Fig. 1. The main two activated pathways refer to ECs (*angiogenesis* and *vasculogenesis*). Angiogenesis refers to formation of new blood vessels generated by existing blood vessels, while vasculogenesis refers to circulating progenitor cells that can get anchored in the existing blood vessels in tissues and form new blood vessels. The latter mechanism occurs in embryonic stages, but it can occur also in wet AMD (Machalinska et al, 2011). It is also possible that vasculogenesis does not occur in the choroid of dry AMD, but it is selected by the software because of a large proportion of genes that participate also in "angiogenesis". The "cell viability of tumor cells" pathway refers, in the context of ChECs not to tumor cells, but to genes that are involved in increased survival, resistance to apoptosis, response to inflammation or oxidative stress, stress resistance, pro-angiogenic survival signaling of ChECs. The upregulated STAT3 gene (orange) is known to significantly promote angiogenesis (Chen and Han, 2008), while the downregulated COL18A1 (collagen type XVIII) (blue), known to generate the anti-angiogenic protein endostatin, is expected to increase angiogenesis (O'Reilly et al, 1997) Therefore all four pathways indicate increased angiogenesis.

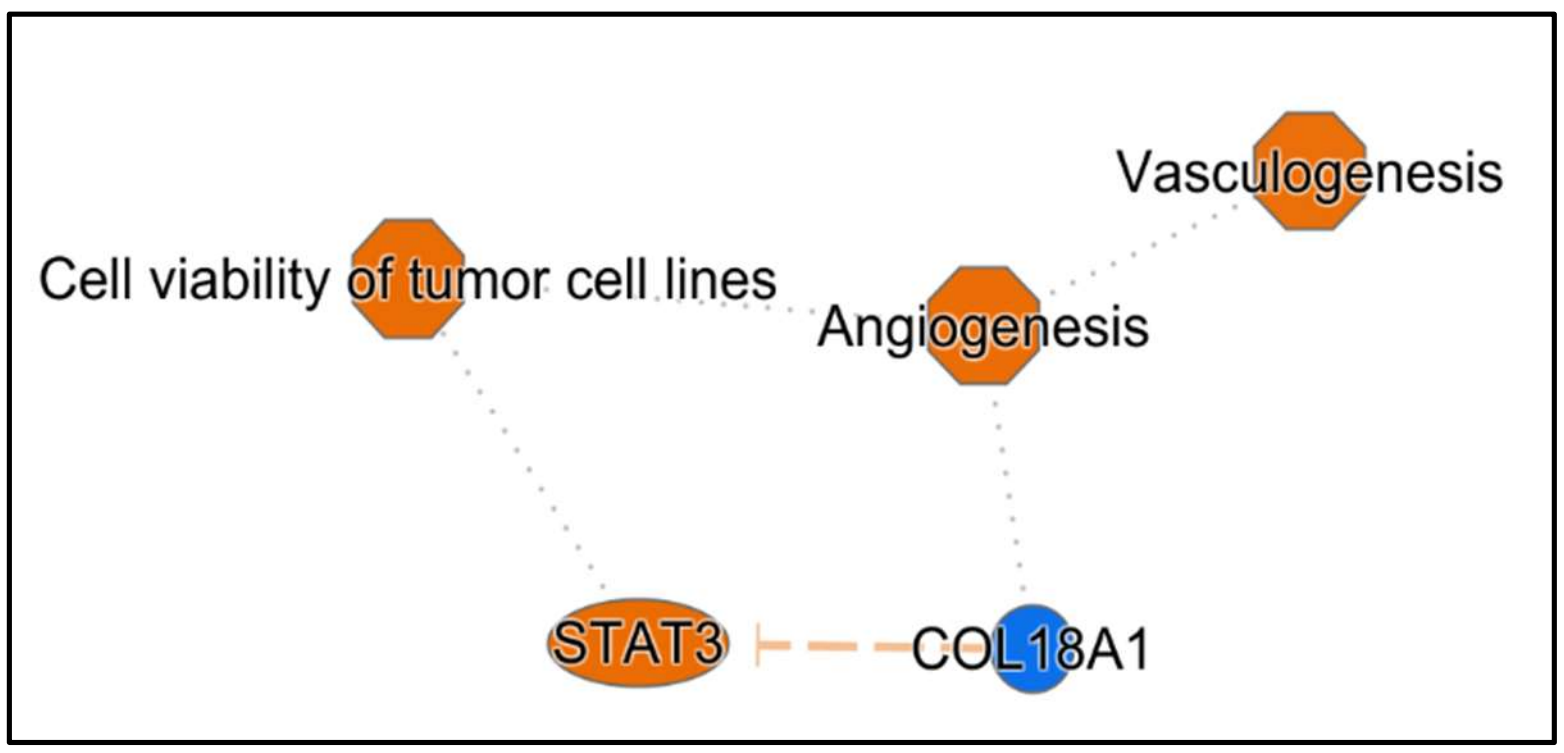

Fig. 1. Graphical summary of the Ingenuity Pathway Analysis (IPA). The Differentially Expresed Genes in dry AMD versus the control subjects are significantly enriched in the five pathways represented in the figure. The octagons correspond to biological functions, and are the higher-level “outcome” nodes, predicted by the software to be activated (orange color). Other pathways or protein complexes involved are STAT3 (upregulated - orange) and COL18A1 (collagen type XVIII) (downregulated - blue). The dashed orange line terminated by a bar indicates indirect inhibition, meaning that intermediary steps between the two partners may exist. The dotted lines represent interactions that are inferred / predicted (literature based). The gray color indicates that no directionality of the interaction can be inferred. As detailed in the text, four of the five entities in the diagram indicate upregulated angiogenesis. Details about known relations among the entities are provided in the text.

*The disease and functions module* (the section cardiovascular diseases) indicates 3 processes as most significantly upregulated. The first is *Angiogenesis* (352 genes in the IPA pathway, 7 out of 10 DEGs upregulated (CCL2, VEGFC, ANXA3, DLL4, CD34, NOTCH4, FOS) out of 10 DEGs, 1 opposite, increased in DEGs and decreased in the pathway, and 2 affected with no direction specified). The second upregulated process is *Vasculogenesis* (6 out of 8 DEG upregulated, (CCL2, VEGFC, DLL4, CD34, RAMP2, FOS), one opposite, and 1 affected). The third upregulated process is *Migration of ECs* (5 of 7 DEGs upregulated (CCL2, PTP4A3, VEGFC, ANXA3, THBD), 1 affected, and one opposite). Although the number of DEGs involved is low (5-7), they represent a larger percentage of the total number of DEGs (11-17.5%).

*The Canonical pathways module* retrieves only 1 significantly enriched pathway, the transcription factor HEY1, upregulated (4 DEGs out of the 157 genes in the pathway: DLL4, VEGFC, NOTCH4, JAG2). SUMOylation of HEY1 represses numerous angiogenic genes (Ren et al., 2024), and therefore, this factor also indicates upregulated angiogenesis, as do the other mentioned pathways.

Dataset 2**.** EC clusters have been obtained from 4 intermediate AMD and 6 controls. Pseudobulk counts for 2191 genes were analyzed by DESeq2, which indicated that 50 genes were significantly downregulated and 34 upregulated.

*The graphical summary* (not shown) consists of only 2 nuclear RNAs, coding for mitochondrial proteins (LONP1 and DAP3), both downregulated, connected by an inferred relation.

*The Diseases and Functions module* (the cardiovascular development section) shows 9 pathways, all referring to ECs; 7 of these contain >=4 DEGs, all upregulated. Among these are: (a) *angiogenesis* (16 DEGs, of which 10 are in the correct direction (7 up-up and 3 down-down), 1 unspecified, and 4 in opposite directions), (b) vasculogenesis (14 DEGs, 9 in correct direction, 1 unspecified), (c) migration (11 DEGs, 9 correct, 1 opposite, 1 unspecified), and (d) development of endothelial tissue (10 DEGs, 6 in the correct direction, 1 affected). The pathways mentioned above include 17 unique DEGs (20% of the 84 DEGs).

*Canonical pathways*: indicate downregulation of mitochondrial processes, (8 out of the top 10 pathways), mostly related to RNA processing. For example, some of these processes are mitochondrial RNA degradation (7/25 DEGs), mitochondrial rRNA processing (7/32), mitochondrial tRNA processing (7/41), but also energy production (oxidative phosphorylation (8/102), and respiratory electron transport (7/89); One pathway is "mitochondrial dysfunction".

Mitochondrial perturbations could be real but could also be caused by the filtering procedure included in the scRNA-seq, which excludes cells that have a high content of mitochondrial genes, indicative of damaged cells. Filtering is not supposed to favor disease vs. control samples, but we cannot exclude the possibility that some non-obvious bias is introduced.

Another upregulated pathway is "*Role of PKR in interferon induction and antiviral response* 7/132, 6 in the same direction, 1 not specified.

*Upstream analysis* factors reveal 6 factors downregulated, and one upregulated. The top downregulated factor is DAP3, which is a mitochondrial ribosomal member (7 DEGs controlled by it, all downregulated) and "Stress granule signaling pathway" (6/247), all 6 down.

**PathfindR** is an R package for active-subnetwork-oriented pathway enrichment analyses (Ulgen et al., 2019).

Dataset 1**.** PathfindR analysis revealed that the KEGG gene database combined with String Protein-Protein Interaction Network (PIN) and KEGG combined with Biogrid PIN are most informative: the first retrieves as the most significant pathway, *Notch signaling* (5 DEGs upregulated). Notch signaling is a cell-to-cell communication pathway that regulates angiogenesis by controlling tip/stalk cell differentiation, proliferation, and migration, and ensuring proper vessel sprouting and patterning (Muñoz-Chápuli et al., 2004). Patterning and branching help establish the correct, stable network of vessels. The 5th most significant pathway is *TNF signaling* (5 DEGs, all activated). This pathway interacts with Notch (Fazio & Ricciardiello, 2016; Jiao et al., 2012; Maniati et al., 2011), and the interaction can be synergistic (both pathways enhance certain outcomes), or antagonistic (TNF-α inhibits Notch signaling). Another pathway that appears in both databases and could be relevant for dry AMD is **"**Fluid shear stress and atherosclerosis" (5 DEGs upregulated), discussed below.

Two pathways could be connected with dry AMD by the presence of calcium deposits**:** "osteoclast differentiation" (5 genes upregulated, FOS, JUNB, SOCS3, CYBA, ATP2A3), and "endocrine resistance" (of hormone-responsive breast cancer) (JAG2, NOTCH4, FOS, DLL4, all upregulated). These pathways have meaningful connections with dry AMD, as discussed later.

Dataset 2. The top pathways retrieved across most databases include 11 mitochondrial pathways, all downregulated. All pathways are based on 6 mitochondrial RNAs, which participate in mitochondrial processes (tRNA processing, RNA degradation, rRNA processing, mitochondrial protein degradation, respiratory electron transport, aerobic respiration). Other downregulated pathways are "cellular response to stress" (7 DEGs downregulated (DNAJA1, HSP90AB1, HSPA8, HSPH1, MT-CO1, MT-CO3, SQSTM1 and "Protein processing in endoplasmic reticulum (4 genes downregulated (HSPA8, DNAJA1, HSP90AB1, HSPH1, are also included in "cellular response to stress").

**ROntoTools.** (Voichita et al., 2025) is a topology-based pathway analysis method that identifies significantly impacted pathways using the entire set of measurements, not only the DEGs, addressing the fact that existing methods fail to distinguish between the primary deregulation of a given gene itself and the effects of signaling coming from upstream.

For the Dataset 1 and the KEGG database, the top 2 pathways are "Notch" and "TNF" (4/62 and 5/119, respectively). Another possibly biologically relevant pathway is "Fluid shear stress and atherosclerosis" (5/152). This pathway refers to ECs in large arteries affected by atherosclerosis, but theoretical models suggest a role in AMD of shear stress in ChECs (Gelfand et al., 2016).

An intriguing output is “osteoclast differentiation” (5/142). The connection between this pathway and dry AMD is detailed later.

For Dataset 2, ROntoTools and the KEGG database yield as pathways possibly relevant for dry AMD “fluid shear stress and atherosclerosis” (6/142), (also found in Dataset 1 by PathfindR and RontoTools), and “protein processing in endoplasmic reticulum” (7/174), found also in Dataset 2 by PathfindR, and “rheumatoid arthritis“ (4/95). RA patients have a higher risk of developing AMD, particularly dry AMD, while some studies suggest RA may even protect against wet AMD (Yoon et al., 2024). RA involves chronic inflammation that activates endothelial cells, leading to endothelial dysfunction (Bordy et al., 2018).

For Dataset 3, ROntoTools did not detect any differential pathways.

**fGSEA** (Fast Gene Set Enrichment Analysis) is an R/Bioconductor package for preranked gene set enrichment (GSEA) (Subramanian et al., 2005), which determines if predefined gene sets (pathways or functional categories) are significantly enriched at the extremes (top/bottom) of a gene list ranked by differential expression (Korotkevich et al., 2021).

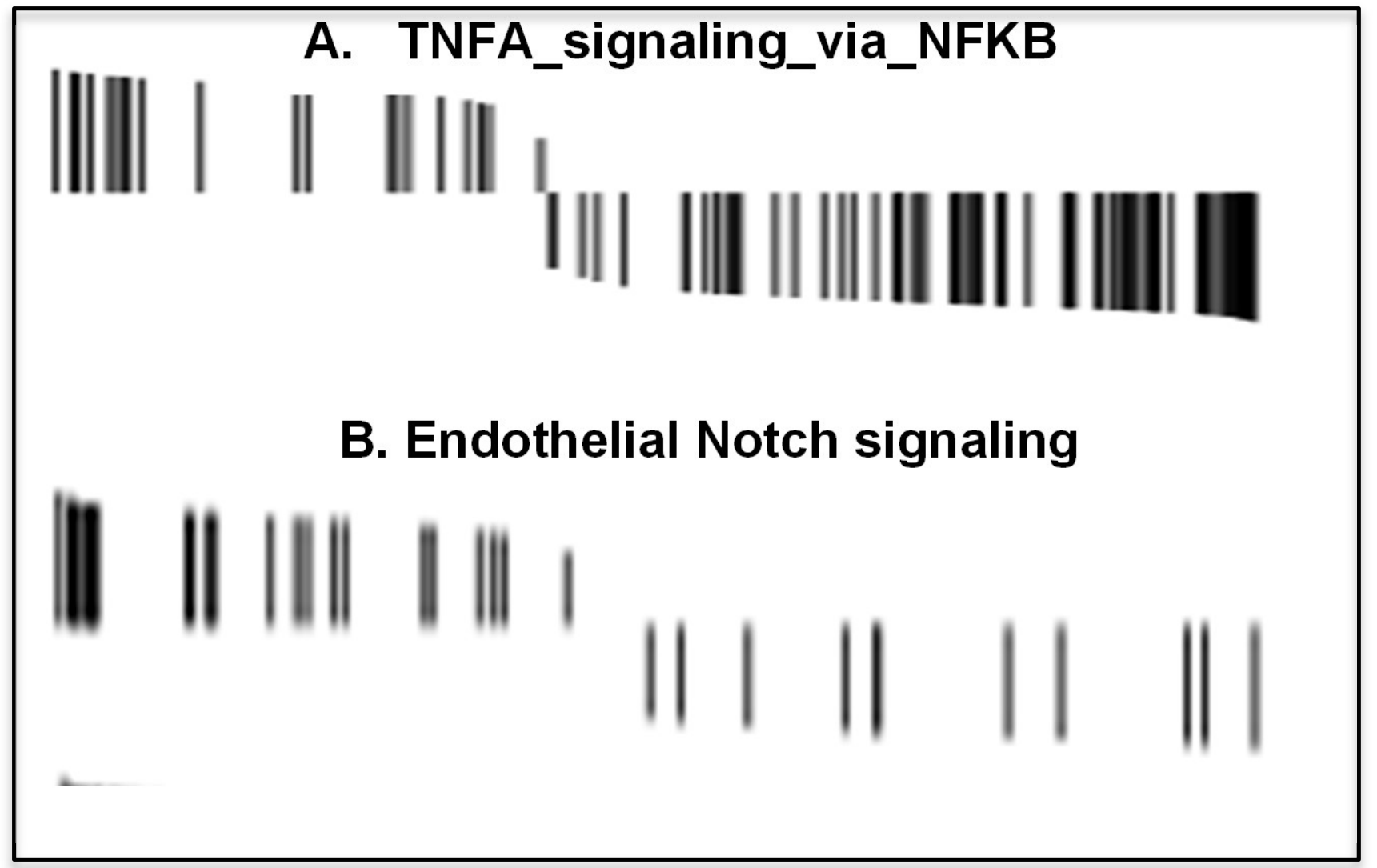

Fig. 2. Representative output of fGSEA package showing enrichment of the Differentially Expressed Genes in Dataset 1 in genes included in two pathways: “TNFA_signaling_via_NFKB” (panel A) and “Notch signaling” (panel B). Each vertical line corresponds to an RNA that belongs to the respective pathway, and is found in the analyzed dataset. The height of each bar is proportional with the increase or decrease in the expression level. The position of each bar on the horizontal axis indicates the position of the RNA species in the list of detected RNAs, which was ranked in decreasing order of differential expression. Higher density of the bars to the left side corresponds to the genes in the pathway that are present with high frequency among the genes upregulated in AMD. Higher bar density to the left corresponds to the genes in the pathway that are downregulated in AMD. (If the RNAs in the pathway would have no relation with the increased or decreased expression in AMD, the bars would be uniformly distributed along the horizontal axis). The two pathways are directly relevant for ECs, as detailed in the text.

For Dataset 1**,** fGSEA retrieves from the Hallmark database, among the highly significant sets, some that are plausibly supporting a role of ECs in dry AMD, or have been retrieved by some of the packages mentioned above: “TNFA signaling via NFKB” (pAdj = 9E-5) (Figure 2 A), "Endothelial Notch signaling“ (Fig. 2B, complement” (pAdj = 0.004), and “hypoxia” (pAdj = 0.004). The first pathway is listed in the dataset Hallmark (Liberzon et al., 2015) as indicating "inflammatory endothelial activation"; the second pathway, "Endothelial notch signaling" is associated with vascular patterning and vessel maturation, especially during angiogenesis (Roca and Adams, 2007).

In the C2 database (curated gene sets), in the top 50 enriched (p-values in the range of 1.4E-6 to 0.01), there are 6 gene sets for hypoxia.

For the C4 database (computational gene sets focused on cancer), the enriched pathways are “Endothelial_NOTCH_signaling” (p = 4E-8)**,** (Fig. 2 B) and four sets about ECs in various tumors (p = 1E-6 to 0.0008),

For the C5 database, (ontology gene sets) possibly relevant sets are "GOBP_calcium_ion transport"(p = 0.018), and "GOBP_calcium_ion_binding"(p = 0.024). Multiple sets refer to blood vessels: "vasculature development", "endothelium development", blood vessel morphogenesis, circulatory system development, vascular processing, circulatory system, "tube development", and "tube morphogenesis" ("tube" refers in this case to both endothelial and epithelial cells (p = 0.018 - 0.05). Another set is "Abnormal_macular_morphology" (p = 0.025).

For Dataset 2, fGSEA retrieves from the C2 database: "Mitochondria_gene_module" (p = 0.002), "TNF_response not_via-P38" (p = 0.017), and "Response to UV_NHEK_UP" (p = 0.023). (The retina is particularly sensitive to UV irradiation. From the C4 database, the sets retrieved are "Endothelial_stress" (p = 0.002) and "Endothelial_NOTCH_signaling" (p = 0.009)

The overall conclusion that emerges from the above analyses is that, although the number of samples is small, comparative analyses done by multiple software packages lead to a coherent and plausible picture. The predominant theme is upregulation of angiogenic signaling, detected by most packages in both early and intermediate AMD. This observation is unexpected and apparently paradoxical. It is unexpected because enhanced angiogenesis is not a feature of dry AMD, but rather the hallmark of wet AMD. The observation is paradoxical because numerous studies, mentioned in the introduction, reported a decrease, rather than an increase, in the number and function of the choroidal vasculature in dry AMD.

A hypothesis that reconciles and explains this apparent contradiction is that in dry AMD, angiogenic processes are triggered by angiogenic "push signals", but the formation of blood vessels is not successfully executed. Moreover, not only that angiogenesis not executed, but the last step in the normal chain of events (angiogenesis activation, followed by execution, followed by deactivation of angiogenesis) is also not finalized and not closed. Therefore, instead of experiencing a transient proangiogenic state, the ECs remain in a non-physiological extended/chronic activation, which could lead to EC damage or death. This mechanism could explain the widely reported decay of the choroidal vasculature in dry AMD. More specifically, the "ghost vessels" (vessel-like channels containing no RBCs and ECs), frequently reported in early-intermediate AMD, could occur due to EC death.

The hypothesis about failed angiogenesis is intellectually appealing because it proposes a unitary mechanism for the two main forms of AMD, dry and wet: both forms are caused by a demand for new blood vessels. The difference is in the response to this demand: in some subjects, the ECs respond by excessive proliferation (wet AMD), while in others, the ECs cannot successfully execute the proliferation program and are destroyed while trying (dry AMD). An alternative explanation may be that while these pro-angiogenic signals derived from the ChECs, Bruch's membrane dissolving signals are being sent from macrophages, RPE, glia or in combination to recruit vasculogenesis from local nearby choroid. As RPE die, and a hole forms in the macula VEGF secreted from RPE, peripheral to the hole, flows into the new sink, which may reach healthier choroid further away than the locally dying choroid. This less damaged choroid receiving the pro-angiogenic signals and preserved growth capacity initiate neovascularization and this is where the CNV in "Wet AMD" receive their aberrant growth.

This mechanism is not unprecedented. A similar process was proposed for another age-related degenerative disease, Alzheimer's disease (AD) (Grammas et al., 2011; Thirumangalakudi et al., 2006). Microvessels isolated from the brains of AD patients express a large number of angiogenic proteins, reflecting the fact that the signaling cascades associated with angiogenesis are upregulated. However, there is no increased brain vascularity in AD. Perpetual EC activation results in the release of a large number of proteases, inflammatory proteins, and other gene products that, after extended abnormal exposure, can injure or kill the ECs and the surrounding cells. This concept was supported by experiments that showed that pharmacologic blockade of vascular activation improved cognitive function in an animal model of AD (Grammas et al., 2011; Thirumangalakudi et al., 2006). Regarding the upstream cause of angiogenesis activation, Hypoxia Inducible Factor 1-alpha (HIF-1α) was found to be elevated in the microcirculation of AD patients. HIF1a is a key regulator of cellular responses to hypoxia, including angiogenesis (Grammas et al., 2011).

Similarities exist between AMD and AD, beyond the fact that both are degenerative age-related diseases. For instance, the vascular network in the eyes of persons who have AD is disorganized. Amyloid β, a key factor in AD, has been repeatedly identified in drusen/sub-RPE deposits, including reports that Aβ deposition was specific to drusen from dry AMD eyes. In that sense, dry AMD and Alzheimer's disease appear to share an amyloid-associated degenerative/inflammatory program (Yoshida et al., 2005; Wang et al., 2021).

Regarding angiogenesis induction in dry AMD, a legitimate question is which are the angiogenic "push" signals. We discuss some candidates below.

*VEGF* (Vascular Endothelial Growth Factor) and SDF-1α (Stromal cell-derived factor 1) are higher in the aqueous humor of intermediate dry AMD vs controls (Keles et al., 2021). VEGF is the most studied angiogenesis inducer, and SDF-1α / CXCL12 is a key chemokine that promotes angiogenesis (Deshane et al., 2007). Higher VEGF was detected in the aqueous humor in eyes with soft drusen and subretinal drusenoid deposits vs. soft drusen only and controls, and correlated with drusen volume (Yoon et al., 2024). VEGF and HuR were increased in the RPE in dry AMD vs controls (Bresciani et al., 2024). In many contexts, HuR can stabilize/boost mRNAs for angiogenesis-related genes, especially VEGFA (Hamm et al., 2026). However, attempts to demonstrate a causal role of VEGF in dry AMD have failed so far: anti-VEGF antibodies delivered preventively in high-risk dry AMD did not show a clear benefit (Heier et al., 2021). Quarterly aflibercept did not lower conversion to exudative AMD, and quarterly ranibizumab likewise did not reduce conversion or improve visual acuity (Chan et al., 2022). SDF1a and HuR remain to be explored in this context.

Another process that could trigger angiogenesis in dry AMD can be the formation of *membrane attack complexes (MAC).* In ECs in culture, complement attack triggers a strong endothelial stress/injury transcriptional program that involves remodeling/angiogenesis-related genes (Zeng et al., 2016).

Another factor that could induce angiogenesis in dry AMD is *smoking.* Smoking induces hypoxia via exposure to carbon monoxide, which binds to hemoglobin instead of oxygen, thus preventing oxygen from being transported to tissues. Hypoxia induces cells to send signals that they lack sufficient oxygen, which in turn triggers angiogenesis. This mechanism could provide an explanation for smoking being a risk factor for AMD. Over time, smoking can impair lung function and gas exchange, which can contribute to more persistent hypoxemia/hypoxia, especially in COPD (Fricker et al., 2018). Interestingly, COPD is associated with dry AMD. A study in a large Taiwan population found that people with COPD had a higher risk of developing AMD overall, and specifically dry AMD (Bair et al., 2021). Another mechanism by which smoking can increase the risk of AMD by triggering angiogenesis is based on Reactive Oxygen Radicals (ROS). Smoking induces oxidative stress in Retinal Pigment Epithelial cells (Cano et al., 2010). It is possible that smoking induces oxidative stress in the ECs as well. Low to moderate ROS in ECs can act as pro-angiogenic signals, promoting pathways like VEGF/VEGFR2 (Kim et al., 2014).

To summarize, previous studies support the possibility that VEGF secretion, MAC formation, and smoking could activate angiogenesis pathways in dry AMD. Two other factors, SDF-1α and HuR, are increased in dry AMD, and their triggering roles remain to be investigated.

A concept relevant for the proposed mechanism of participation of ECs in dry AMD is *"endothelial activation"* (Al-Soudi et al., 2017). The basic idea is that ECs are turned "On", meaning that they leave their normal state, in which they provide the stable structures that allow blood flow. EC activation can be of two types, inflammatory and angiogenic.

In *inflammatory activation*, ECs become “sticky” and leaky to recruit immune cells. Markers for inflammatory endothelial activation include adhesion molecules (ICAM-1, VCAM-1, E-selectin), von Willebrand factor (VWF), and P-selectin, which facilitate leukocyte adhesion and thrombosis. Evidence for inflammatory activation of ECs in dry AMD is very limited: SNPs for SELE, SELL, and SELP have been investigated in dry AMD. A single SNP located within an intron of SELP was statistically associated with dry AMD (Mullins et al., 2011). An in vitro study showed that choriocapillaris ECs in organ culture responded to complement C5a by increasing ICAM-1 mRNA and protein (Skeie et al., 2010).

In the other form of activation — *angiogenic* — ECs switch to a “growth mode” in which they respond to angiogenic factors by proliferation, migration, sprouting, and secretion of Matrix Metallo-Proteinases (MMP), to remodel the matrix and provide space for new blood vessels (Munoz-Chapuli et al., 2004). The ECs also generate tip cells and stalk cells to form new capillary branches. Within the hypothesis on angiogenesis failure in dry AMD, the relevant mechanistic question is whether some of these processes fail. The question remains to be addressed experimentally. A related question is whether some of these processes are activated and remain chronically active, instead of being executed, finalized, and closed.

Besides Alzheimer’s disease, our analysis suggests that EC-mediated mechanisms in dry AMD are also present in rheumatoid arthritis (RA) and in osteoporosis.

Regarding RA, RNAs involved in this disease are enriched among the DEGs found by us in dry AMD. RA involves chronic inflammation that activates EC, leading to endothelial dysfunction (Bordy et al., 2018). Moreover, RA patients have a higher risk of developing dry AMD, while having a lower risk of wet AMD (Yoon et al., 2024). These observations are compatible with the idea that angiogenesis is defective/underactive in dry AMD, and overactive in wet AMD.

Regarding osteoporosis, the connection with dry AMD is via calcification processes. On one side, in dry AMD, bone-like hydroxyapatite (HAP) deposits have been reported. Small spheres of HAP filled with cholesterol, called spherules, were found only in drusen from people with dry AMD, and not in those with wet AMD or without AMD (Rajapakse et al., 2020). On the other hand, DEGs found by our analysis are enriched in the pathway “osteoclast differentiation”. Osteoclasts, the bone-resorbing cells, are a key factor in osteoporosis.

Interestingly, osteoporosis is a risk factor for AMD. The prominent feature of osteoporosis is loss of calcium in bones, but a separate, often concomitant process, occurs in some patients: vascular calcification (VC). VC consists of calcium buildup in the extracellular matrix of the vessel walls. The deposits contain HAP, similar to drusen. ECs can contribute directly to VC: activated ECs can generate osteoprogenitor cells through the process known as endothelial-mesenchymal transitions (EndoMT). This process leads to osteogenic differentiation of the ECs, which in turn can lead to VC (Jiang et al., 2022). Even more interestingly, there is yet another piece of the same puzzle that falls into place: as mentioned, activation of shear stress was detected by pathway analysis, and abnormal shear stress induces EndoMT (Mahmoud et al., 2017).

ECs also secrete pro-osteogenic growth factors, such as bone morphogenetic proteins, inflammation mediators, and cytokines (Zhang et al., 2021). Another related fact is that osteopontin, a bone protein, is a component of AMD basal deposits, and its local expression correlates with calcified, large drusen, which is a hallmark of dry AMD (Lekwuwa et al., 2022). Yet another calcification-related observation is that some of the DEGs detected by us are associated with the pathway “Hormone Receptor negative (hormone resistant) breast cancers”. HAP microcalcifications are commonly associated with malignant breast lesions, and more aggressive calcification patterns have been linked to hormone-resistant breast tumors (O’Grady, 2018).

These observations support the hypothesis that the formation of calcium deposits in rheumatoid arthritis, osteoporosis, and some tumors share common mechanisms with dry AMD, and dysfunctional ECs could be either the key contributors or common sufferers.

Besides EC-specific perturbations the pathway analysis suggests mitochondrial impairment in intermediate AMD. This observation is in line with numerous studies that reported mitochondrial perturbations in dry AMD (Dohl et al., 2025; Luján et al., 2022).

In conclusion, our analysis provided further support to studies that revealed deficiencies in the number and functionality of choroidal blood vessels in dry AMD. The main outcome of our bioinformatics meta-analysis is that the deficiencies in the choroidal vasculature in dry AMD appear to be caused by an unproductive or possibly counter-productive attempt to generate more blood vessels by angiogenesis. Moreover, the dysfunction of choroidal endothelial cells occurs at the initiating stages of geographic atrophy, suggesting that those changes precede retina degeneration and vision loss. From this perspective, dry and wet AMD represent “two sides of the same coin”: in dry AMD choroidal angiogenesis is failing to support the retina, while in wet AMD vision is impaired by excessive angiogenesis.